\documentclass[pre,twocolumn,superscriptaddress]{revtex4-1}
\usepackage{graphicx}
\usepackage{amsmath,amssymb}

\newcommand{\myav}[1]{\langle #1\rangle}

\newcommand{\myvec}[1]{{\mathbf{#1}}}
\newcommand{\rvec}{\myvec{r}}
\newcommand{\kvec}{\myvec{k}}

\newcommand{\kB}{k_\mathrm{B}}
\newcommand{\kT}{\kB T}
\newcommand{\lB}{l_{\mathrm{B}}} 
\newcommand{\kD}{k_{\mathrm{D}}}
\newcommand{\lD}{\lambda_{\mathrm{D}}}
\newcommand{\lRPA}{\lambda_{\mathrm{RPA}}}
\newcommand{\lHNC}{\lambda_{\mathrm{HNC}}}
\newcommand{\qD}{q_{\mathrm{D}}}
\newcommand{\lr}{^{\mathrm{L}}}
\newcommand{\sr}{^{\mathrm{S}}}

\newcommand{\pex}{p^{\mathrm{ex}}}

\newcommand{\rhoz}{\rho_z}

\newcommand{\mol}{\mathrm{mol}}
\newcommand{\metre}{\mathrm{m}}
\newcommand{\nm}{\mathrm{nm}}
\newcommand{\M}{\mathrm{M}}

\newcommand{\latin}[1]{{\itshape #1}}
\newcommand{\eg}{\latin{e.\,g.}}
\newcommand{\ie}{\latin{i.\,e.}}
\newcommand{\etal}{\latin{et al.}}
\newcommand{\etc}{\latin{et\,c.}}

\DeclareMathOperator{\erf}{erf}

\begin{document}

\title{Screening properties of Gaussian electrolyte models, \\
with application to dissipative particle dynamics}

\author{Patrick B. Warren}

\email{patrick.warren@unilever.com}

\affiliation{Unilever R\&D Port Sunlight, Quarry Road East, Bebington,
  Wirral, CH63 3JW, UK.}

\author{Andrey Vlasov}

\affiliation{Department of Chemistry, St. Petersburg State University,
  26 Universitetsky prosp., 198504 St. Petersburg, Russia.}

\author{Lucian Anton}

\altaffiliation{Current address: Computational Science and Engineering
  Department, Science and Technology Facilities Council, Daresbury
  Laboratory, Daresbury Science and Innovation Campus, Warrington,
  Cheshire, WA4 4AD, UK.}

\affiliation{School of Chemical Engineering and Analytical Science,
  University of Manchester, Manchester, M13 9PL, UK.}

\author{Andrew J. Masters}

\affiliation{School of Chemical Engineering and Analytical Science,
  University of Manchester, Manchester, M13 9PL, UK.}

\date{March 4, 2013}

\begin{abstract}
We investigate the screening properties of Gaussian charge models of
electrolyte solutions by analysing the asymptotic behaviour of the
pair distribution functions.  We use a combination of Monte-Carlo
simulations with the hyper-netted chain integral equation closure, and
the random phase approximation, to establish the conditions under
which a screening length is well defined and the extent to which it
matches the expected Debye length.  For practical applications, for
example in dissipative particle dynamics, we are able to summarise our
results in succinct rules-of-thumb which can be used for mesoscale
modeling of electrolyte solutions.  We thereby establish a solid
foundation for future work, such as the systematic incorporation of
specific ion effects.
\end{abstract}

\maketitle

\section{Introduction}
Dissipative particle dynamics (DPD) 
has seen widespread uptake in modelling soft condensed
matter \cite{FS02, NMW03}. The attractions are obvious: by coarse
graining over the atomistic degrees of freedom one can access the
relevant length and time scales with only modest computing
requirements.  Polymer phase behaviour \cite{GW97, GM98}, polymer
dynamics \cite{Spe00}, polymer rheology \cite{SvdM+10}, surfactant
mesophase formation kinetics \cite{PWM02}, the properties of
amphiphilic bilayers \cite{GR01, SL02}, and the properties of colloidal
suspensions \cite{BCL+97} have all been investigated by the method.

Charged systems such as anionic and cationic surfactants,
water-soluble polyelectrolytes, charge-stabilised colloidal
suspensions, and mixtures of these \cite{Gro03}, form a large subclass
of widespread practical importance.  In these systems there is often
the requirement to model the supporting electrolyte.  This can be done
implicitly, for example with the Poisson-Boltzmann equation, or
explictly by incorporating ions as charged particles in the
simulation.  In the latter case, particularly for DPD where soft
interactions are the norm, it is natural to smear the point charges
into charge clouds.  The divergence of the long-range Coulomb law as
$r\to0$ (where $r$ is the center-center separation) is replaced by a
smooth cutoff, thus ensuring thermodynamic stability according to a
theorem by Fisher and Ruelle \cite{MEF66}.

The precise form of the charge smearing is often tuned to the choice
of numerical algorithm and a consensus on the best approach has yet to
emerge.  Groot introduced a grid-based method with linear charge
smearing \cite{Gro03}. Later Gonz\'alez-Melchor \etal\ examined an
Ewald-based method with exponential charge smearing \cite{GMV+06}. Here
we study a related Ewald method with Gaussian charge smearing.  This
choice can be used to simplify the Ewald algorithm, and connects with
recent work on the so-called ultrasoft restricted primitive model
(URPM) \cite{CHK11a, CHK11b, NHK12}. In principle the differences
between smearing methods can be subsumed into short-range part of the
interparticle potential, though the details are the subject of ongoing
investigations.

To study the screening properties of our Gaussian electrolyte
model, we use a combination of Monte-Carlo (MC) simulations, the
hyper-netted chain (HNC) integral equation closure, and the random
phase approximation (RPA), to analyse the asymptotic behaviour of the
pair distribution functions.  The programme is as follows.  In the
next two sections we define the mesoscale electrolyte model and the
tools used to analyse it.  We then present results demonstrating that,
for typical applications, HNC can be relied upon to deliver accurate
results (it is no exaggeration to say that HNC is up to ten million
times faster than MC).  We then use HNC to explore the screening
properties of the model, establishing the conditions under which a
screening length is well defined (\ie\ on the low density side of a
Kirkwood line in the phase diagram) and the extent to which the
screening length matches the expected Debye length.  We further
establish the domain of applicability of the much simpler RPA, which
gives relatively simple expressions for the Kirkwood line and the
screening length.  We emphasise that our approach could easily be
applied to other smeared charge electrolyte models.  Mindful of this,
and the utility of a fast, accurate, multicomponent HNC solver in
general, we have released our FORTRAN 90 HNC code as fully documented
open source software \cite{OZnote}.

\section{Model}
\label{sec:model}
We now describe the Gaussian charge model for electrolyte solutions.
The potential energy is given by a sum of pairwise terms, split into
short range and long range (electrostatic) contributions,
\begin{equation}
U=\sum_{i>j} U_{ij},\quad U_{ij}=U\sr_{ij}+U\lr_{ij}\,.
\end{equation}
The short range piece is given by 
\begin{equation}
\beta U\sr_{ij}=\left\{\begin{array}{ll}
\frac{1}{2}A_{ij}(1-r_{ij}/r_c)^2 & (r_{ij}<r_c)\\[3pt]
0 & (r_{ij} \ge r_c)
\end{array}\right.
\label{eq:usrij}
\end{equation}
and the long range piece is given by
\begin{equation}
\beta U\lr_{ij}=
\frac{\lB z_i z_j}{r_{ij}}\erf\Bigl(\frac{r_{ij}}{2\sigma}\Bigr)\,.
\label{eq:ulrij}
\end{equation}
In these $\beta=1/\kT$ is the inverse of the temperature $T$ measured
in units of Boltzmann's constant $\kB$, $r_{ij}$ is the centre-centre
separation between particles $i$ and $j$, $A_{ij}$ is a dimensionless
short range repulsion amplitude which depends on the particle types,
$\lB$ is the Bjerrum length which plays the role of an electrostatic
coupling constant, $z_i$ and $z_j$ are the valencies measured in units
of an elementary charge, and $r_c$ and $\sigma$ are length scales
which measure, respectively, the range of short range repulsion and
the size of the Gaussian charge cloud.  The short range part of the
potential corresponds to the standard DPD interaction law \cite{GW97}.
The long range part corresponds to the interaction between Gaussian
smeared charges with a radial charge distribution
$(2\pi\sigma^2)^{-3/2} \exp({-r^2/2\sigma^2})$.  The function
$\erf(r/2\sigma) \to r/(\sigma\sqrt{\pi})$ as $r\to0$, thus ensuring
the Coulombic divergence is replaced by a smooth cutoff.

We will consider up to three species of particles, corresponding to
positively and negatively charged ions of valencies $z_+$ and $z_-$ at
densities $\rho_+$ and $\rho_-$, and a third neutral solvent species
at a density $\rho_0$.  The total \emph{ion} density will be denoted
by $\rhoz=\rho_++\rho_-$.  The total \emph{overall} density will be
denoted by $\rho=\rho_0+\rhoz$.  In the case where there is no
solvent, $\rho_0=0$ and $\rho=\rhoz$.  We adopt the convention that
the valency includes the sign as well as the magnitude. Overall charge
neutrality then requires $z_+\rho_++z_-\rho_-=0$.  We do not
necessarily suppose the valencies are of the same magnitude.  We shall
label species by Greek indices, $\alpha, \beta = (0, +, -)$.  The
system volume is $V$ and thermal averages will be denoted by
$\myav{\cdot}$.

We first consider a special case.  The aforementioned URPM is an
unsolvated equimolar mixture of Gaussian charge clouds, corresponding
to the choice $A_{ij}=0$, $\rho_0=0$, and $|z_\pm|=1$.  The URPM is
governed by a dimensionless density, $\rhoz\sigma^3$, and a
dimensionless coupling constant, $\lB/\sigma$, which plays the role of
an inverse temperature.  The model exhibits marked clustering for
$\lB/\sigma\agt30$, and a condensation transition for
$\lB/\sigma\agt100$, for densities in the range
$\rhoz\sigma^3\approx0.01$--0.1 (these estimates are translated from
the results shown in Fig.~5 in Ref.~\onlinecite{CHK11a}).  The physics
behind the phase transition remains somewhat unclear \cite{NHK12,
  WM13}, but the phenomenology can be viewed as a reflection of
stability issue for point charges mentioned in the introduction.  It
quantifies the onset of the `danger zone' as the point charge limit is
approached.  To avoid these artefacts the implication is that we
should attempt to keep $\lB/\sigma\alt30$.  However for practical
applications there is already a strong incentive to make $\sigma$ as
large as possible, to reduce the cost of computing the electrostatic
interactions.  Usually this is enough to ensure that low temperature
URPM artefacts are avoided.

In the general case the properties of the model are
governed by three length scales, $r_c$, $\sigma$ and $\lB$, the
repulsion amplitude matrix $A_{ij}$, the choice of valencies
$z_{\pm}$, and the densities $\rhoz$ and $\rho$.  The parameter space
is thus potentially very large.  Our strategy to reduce the complexity is
to consider the mapping to the underlying atomistic system.
This requires us to distinguish between \emph{physical} units in which
the length scales and densities are expressed in SI units; and
\emph{simulation} units in which length scales and densities are
expressed in units of $r_c$ or $\sigma$.

In standard DPD the choice $\rho r_c^3=3$ is usually made \cite{GW97},
and we will adopt the same here.  In addition one usually introduces
the notion of a `mapping number' $N_m$, giving the number of solvent
molecules represented by one DPD solvent particle.  Given this, the
value of $r_c$ in physical units is determined by the identity $\rho
N_m V_m/N_A\equiv1$, where $V_m$ is the solvent molar volume and $N_A$ is
Avogadro's number \cite{GMV+06}. If water is the solvent
($V_m=18\times10^{-6}\,\mol\,\metre^{-3}$), and with the conventional
choice $N_m=3$, one has in physical units $r_c= 0.645\,\nm$.

Next consider the Bjerrum length.  In physical units this is
$\lB=e^2/(4\pi\epsilon_r\epsilon_0\kB T)$ where $e$ is the elementary
charge, $\epsilon_r$ is the relative permittivity of the solvent, and
$\epsilon_0$ is the permittivity of free space.  For water at room
temperature $\lB\approx0.7\,\nm\approx 1.09\, r_c$.  Since a $z$:$z$
electrolyte is equivalent to a 1:1 electrolyte with $\lB$ increased by
a factor $z^2$, there is considerable interest in exploring higher
values of $\lB$.  In the present work we shall explore up to $\lB=10\,
r_c$ (${}\approx 7\,\nm$ in physical units), which covers many cases
of interest.  

For an electrolyte at a molar concentration $c_s$, the microscopic 
ion density is $10^3c_sN_A$.  This is readily converted to a simulation
density by multiplying by $r_c^3$.  For example a $0.1\,\M$  1:1
electrolyte solution would be represented by $\rhoz r_c^3=0.032$ 
(note that $\rho_z$ counts both species of ion).  To
cover the typical range of electrolyte concentrations we therefore
consider $\rhoz r_c^3$ in the range $10^{-3}$--1.

The above considerations do not yet impinge on the
choice of $\sigma$.  This is a central theme of the present study and
will be discussed extensively below.

Finally we discuss the repulsion amplitude matrix.  In the present
study we will only consider a constant repulsion amplitude matrix
$A_{ij}=A$, leaving the extension to unequal repulsion amplitudes
for future work. We use either $A=25$ motivated by standard
DPD \cite{GW97}, or $A=0$ corresponding to the situation in the absence
of short range repulsions.  In the latter case, of course, it does not
make sense to include the neutral solvent species since it would just
form an ideal gas in the background.  It has been suggested that $A$
should be chosen to match the solvent compressibility \cite{Gro03}.
However this introduces the danger if $A$ is too large one will
encounter an order-disorder transition driven by the short range
repulsions.  A less rigorous criterion is to demand only that the
solvent be relatively incompressible, so that $\partial(\beta
p)/\partial\rho\gg1$ where $p$ is the pressure.  This is satisfied by
$A=25$, for which the solvent is clearly still a liquid with only
moderate structure.  Moreover much work has been done based on this
value, which we will therefore continue to use.

To summarise, in the remainder of this work we shall consider mainly
two classes of models: either the pure URPM comprising unsolvated
Gaussian charges, or the `solvated' case containing in addition a
neutral species and short range repulsions between all particles.  (It
is possible to consider an intermediate case where short range
repulsions are added to the URPM, however this does not generate any
new insights.)  As already mentioned the URPM is characterised by the
dimensionless density $\rho_z\sigma^3$ (with $|z_\pm|=1$ implying
$\rho_\pm=\rho_z/2$) and coupling strength $\lB/\sigma$.  The length
scale $r_c$ plays no role, except, perhaps, as a `fiducial' length.
On the other hand the solvated case (with the `standard' choice $\rho
r_c^3=3$ and $A=25$) is characterised by the dimensionless densities
$\rho_\pm r_c^3$, and dimensionless ratios $\lB/r_c$ and $\sigma/r_c$,
where all except $\sigma/r_c$ are fixed by the mapping to the
underlying physical system.

From a practical point of view $\sigma\ne r_c$ is commonplace, and it
shall be important to pay attention to the units of length when
mapping between the solvated case and the pure URPM.  In the text we
shall endeavour to be always explicit about this, and where the figure
annotations use implicit units we shall always state the choice of
units in the caption.  A symmetric $z$:$z$ electrolyte in the solvated
case can be mapped to the URPM with a renormalised $\lB\to z^2\lB$.
An asymmetric electrolyte cannot be mapped onto the URPM but we shall
discuss this case only rather briefly.  It is worth bearing in mind
that it is quite straightforward to apply the tools developed here to
\emph{all} these cases.

\section{Tools}
\subsection{Pair distribution functions and screening}
Given the model is governed solely by pair interactions, the
thermodynamic properties are completely determined by the pair
distribution functions $g_{\alpha\beta}(r)$.  In addition the
screening properties are also determined by the asymptotic behaviour
of these functions.  Specifically, the screening length $\lambda$
features in the asymptotic behaviour of the total correlation
functions,
\begin{equation}
h_{\alpha\beta}(r) \equiv g_{\alpha\beta}(r) - 1
\sim \frac{e^{-r/\lambda}}{r}\quad (r\to\infty)
\label{eq:ldef}
\end{equation}
\emph{provided} the asymptotic decay is purely exponential.  The decay
length defined in this way is unique to each state point and does not
depend on the identity of the species of charged particles under
consideration.

As the density decreases the screening length approaches the
Debye-H\"uckel limiting law behaviour, $\lambda\to\lD$, where the
Debye length is
\begin{equation}
\lD=({\textstyle4\pi\lB\sum_\alpha\! z_\alpha^2\rho_\alpha})^{-1/2}\,.
\label{eq:ld}
\end{equation}
Conversely, as the density increases the screening length gets smaller
but there comes a point where the asymptotic decay of the total
correlation functions ceases to be purely exponential and instead
becomes damped oscillatory.  This transition defines a line in the
phase diagram known in charged systems as the Kirkwood line \cite{Kir36}, 
or more generally a Fisher-Widom line \cite{FW69}.

For applications, one would hope that the actual screening length will
hew as closely as possible to the expected Debye length (at least, as
long as the latter is well defined).  The extent to which this can be
made so is the central theme of the present work.  For example, a 1:1
electrolyte at $0.1\,\M$ concentration has a Debye length $\lD\approx
0.96\,\nm\approx1.5\, r_c$.  If we simulate this in the present
model with the choice $\sigma=r_c$, the asymptotic decay of the total
correlation functions would be \emph{oscillatory} and we would not even
be on the right side of the Kirkwood line.  On the other hand if we
use $\sigma=r_c/2$ the asymptotic decay would be purely
exponential with $\lambda/\lD=0.94$, thus the actual screening length
would be only 6\% different from the Debye length.  This example will
be worked through in more detail below.

\subsection{Integral equation theory}
\label{sec:ie}
Given that the underlying electrolyte model is a fluid mixture, it is
natural to think of using multicomponent integral equation theory to
calculate the structural and thermodynamic properties \cite{HM06,
  Ng74, KMP04, VLK+09}. Further, since the interactions are soft, one
expects that the hyper-netted chain (HNC) integral equation closure
should work well.  We find this is indeed the case.  We also find that
for parameters typical of 1:1 electrolytes, the random phase
approximation (RPA) also works well.

The starting point is the multicomponent Ornstein-Zernike (OZ)
relation which defines the direct correlation functions
$c_{\alpha\beta}(r)$.  In reciprocal space the OZ relation is
\begin{equation}
{\tilde h}_{\alpha\beta} = {\tilde c}_{\alpha\beta}+{\textstyle\sum_\gamma}
\,\rho_\gamma\,{\tilde c}_{\alpha\gamma}\,{\tilde h}_{\gamma\beta}
\label{eq:oz}
\end{equation}
where the spatial Fourier transform of a function $x(r)$ is defined by
$\tilde x(k) = \int\!d^3\rvec\, e^{-i\kvec\cdot\rvec} \,x(r)$.
The HNC closure is defined in real space, and is
\begin{equation}
h_{\alpha\beta}=\exp(-\beta
U_{\alpha\beta}+h_{\alpha\beta}-c_{\alpha\beta})-1
\label{eq:hnc1a}
\end{equation}
where $U_{\alpha\beta}$ is the pair potential between particles of
species $\alpha$ and $\beta$.  The solution of these coupled equations
is numerically quite demanding and in the present case we exploit the
accelerated convergence schemes originally proposed by Ng \cite{Ng74,
  KMP04, VLK+09}. We typically solve the distribution functions on a
grid of size 4096 points at a grid spacing $0.01\,r_c$, so that the
functions are calculated out to $r\approx40\,r_c$ where all trace of
structure has typically vanished below the numerical precision of the
calculation.  We find, however, that the schemes fail to converge for
$\lB/\sigma\agt10$.  This loss of solution has also been observed by
Coslovich, Hansen and Kahl \cite{CHK11b}, and may be indicative of a
mathematical property of the HNC rather than a numerical problem.

Numerically much less demanding is the RPA closure, which is given by 
\begin{equation}
c_{\alpha\beta}=-\beta U_{\alpha\beta}\,.
\end{equation}
Because there are no hard cores, the RPA is the same as the mean
spherical approximation (MSA).  Unlike the HNC, the RPA can be
solved for all values of $\lB/\sigma$ although it may yield
unphysical results (for example $h_{\alpha\beta}<-1$).

The pressure $p$ and the internal energy density $\myav{U}/V$, can be
solved from the pair functions \cite{HM06, VLK+09}. The pressure can be
found either by the virial or compressibility routes.  These do not
give exactly the same result because the HNC closure breaks
thermodynamic consistency.  In practice, for the present applications,
we have found the two routes differ by typically at most a few
percent.  Specific results for the RPA thermodynamics can be found in
Refs.~\onlinecite{NHK12} and~\onlinecite{WM13}.

\subsection{RPA solution of the URPM}
\label{sec:rpasol}
Coslovich, Hansen and Kahl solve the RPA for the URPM \cite{CHK11b},
and we have recently revisited the problem in terms of the low
temperature phase behaviour \cite{WM13}. The relevant properties of the RPA
solution are described here.  URPM symmetry implies that in the
RPA the correlation functions are given by $h_{\pm\pm}(r)=\pm h(r)$.
Inserting this into the OZ equations, with the RPA closure,
reveals
\begin{equation}
\tilde h(k)=\frac{-4\pi\lB \exp({-\sigma^2k^2})}
{k^2+\kD^2  \exp({-\sigma^2k^2})}
\label{eq:hk}
\end{equation}
where $\kD^2= 4\pi\lB\rhoz$ is the square of the Debye wavevector
(\ie\ $\kD=1/\lD$).  

The real space total correlation functions can (in principle) be
obtained by expressing the Fourier back-transform of Eq.~\eqref{eq:hk}
as a contour integral in the complex $k$-plane.  The behaviour of the
correlation functions is therefore determined by the poles of $\tilde
h(k)$ in the upper half plane.  As a particular consequence, the
asymptotic behaviour of $h(r)$ as $r\to\infty$ is determined by the
position(s) of the pole(s) closest to the real axis \cite{HAE06,
  NHK12}. There are two cases.  If the nearest pole to the real axis is
purely imaginary, the asymptotic behaviour of $h(r)$ is purely
exponential, with a decay length set by the distance of the pole from
the real axis.  Alternatively if the nearest poles to the real axis are
complex, the asymptotic behaviour is damped oscillatory.  Clearly,
then, the Kirkwood line is determined by the crossover between these
two scenarios, typically when a purely imaginary pole nearest to the
real axis collides with the next-nearest pole, to form a complex pair
which subsequently move off the imaginary axis.

In the present case, writing $q=\sigma k$ and $\qD=\sigma\kD$, the
poles of $\tilde h(q)$ are determined by the solutions of
\begin{equation}
q^2\exp(q^2)=-\qD^2\equiv-4\pi\lB\rhoz\sigma^2\,.  
\end{equation}
These solutions can be expressed in terms of the Lambert $W$ function
which solves $W e^{W}=z$.  For the asymptotic behaviour of $h(r)$ the
most relevant solution is given by $q^2=W_0(-\qD^2)$ where $W_0$ is
the principal branch of the Lambert $W$ function \cite{Cor96, Wnote}.
If $\qD^2\le1/e$, $W_0(-\qD^2)$ is negative real and the corresponding
poles of Eq.~\eqref{eq:hk} (in the complex $q$-plane) are at $q=\pm i
|W_0(-\qD^2)|^{1/2}$.  The corresponding decay length is given by
\begin{equation}
\lRPA=\sigma\times |W_0(-4\pi\lB\rhoz\sigma^2)|^{-1/2}\,.
\label{eq:lrpa}
\end{equation}
Note that $\lRPA\to\lD$ (from below) as $\rhoz\to0$.  If $\qD^2>1/e$,
$W_0(-\qD^2)$ is complex and the asymptotic decay of $h(r)$ is damped
oscillatory.  We therefore identify $\qD^2=1/e$ as the Kirkwood
line \cite{NHK12}. Equivalently, Eq.~\eqref{eq:lrpa} requires
\begin{equation}
4\pi e \lB\rhoz\sigma^2\le1\,,
\label{eq:klrpa}
\end{equation}
with equality determining the location of the RPA Kirkwood line.  For
general applications the charge density in Eqs.~\eqref{eq:lrpa} and
\eqref{eq:klrpa} should be taken to be the ionic strength, defined by
$\rho_z=\sum_\alpha \!z_\alpha^2\,\rho_\alpha$.
The RPA can of course be solved for the solvated URPM case.  We leave 
discussion of this to a separate publication.

\subsection{Monte-Carlo methods}
We benchmark HNC against MC simulations.  We use an $NVT$ ensemble
with standard single particle trial displacements in the usual
Metropolis scheme \cite{FS02}. We calculate the energy of a
configuration from $U=U\sr+U\lr$, where $U\sr=\sum_{i>j}U\sr_{ij}$
from Eq.~\eqref{eq:usrij}, and $U\lr=\sum_{i>j}U\lr_{ij}$ from
Eq.~\eqref{eq:ulrij}.  The latter is re-expressed as an Ewald
sum,
\begin{equation}
\beta U\lr=\frac{2\pi\lB}{V}\! \sum_{k\le k_c} \!\! A_k\,|Q_\kvec|^2
- \frac{1}{2\sigma\sqrt{\pi}}\sum_i z_i^2
\label{eq:uewald}
\end{equation}
where $A_k=k^{-2}\exp({-\sigma^2k^2})$, and $Q_\kvec=\sum_i z_i
e^{-i\kvec\cdot\rvec_i}$ is the reciprocal space charge density.  This
is just the standard Ewald result omitting the real space
contribution \cite{AT87, FS02}. The last term is a self energy
correction; this can of course be omitted from the MC acceptance
criterion, but it is essential to retain this correction when making
comparisons with integral equation theory.

The first term in Eq.~\eqref{eq:uewald} is a sum over a discrete set
of wavevectors, commensurate with the simulation box dimensions, such
that $k=|\kvec|\le k_c$ where the cut-off is chosen so that
$\exp({-\sigma^2k_c^2})$ is sufficiently small.  For the simulations
reported below we use $\sigma k_c=4$.  There are about $4\pi
k_c^3/3\times (2\pi)^3/V$ discrete wavevectors in the sum (the second
factor is the density of wavevectors in reciprocal space).  This means
that the computational cost of evaluating the sum varies as
$1/\sigma^3$.  As a practical consideration, this is a strong
motivation for making $\sigma$ as large as possible.

The only other point to make about the Ewald implementation concerns
the calculation of the pressure, 
\begin{equation}
\begin{array}{l}
\displaystyle
\beta p= \rho-\frac{1}{3V}\Bigl\langle\sum_{i>j} r_{ij}
\frac{\partial(\beta U\sr_{ij}}{\partial r_{ij}}\Bigr\rangle\\[15pt]
\displaystyle
{}\hspace{3em}+\frac{2\pi\lB}{3V^2}\Bigl\langle\sum_{k\le k_c} \!\!
A_k\,|Q_\kvec|^2\,(1-2\sigma^2 k^2)\Bigr\rangle\,.
\end{array}
\end{equation}
The first and second terms in this are the ideal gas result and the
standard virial result for pair interactions.  The third term follows
from Eq.~\eqref{eq:uewald} by calculating $-\myav{\partial
  U\lr/\partial V}$ \cite{Smith87}. Obviously the form of this term
means that it can be easily evaluated alongside the energy.

\begin{figure}
\begin{center}
\includegraphics[clip=true,width=3.0in]{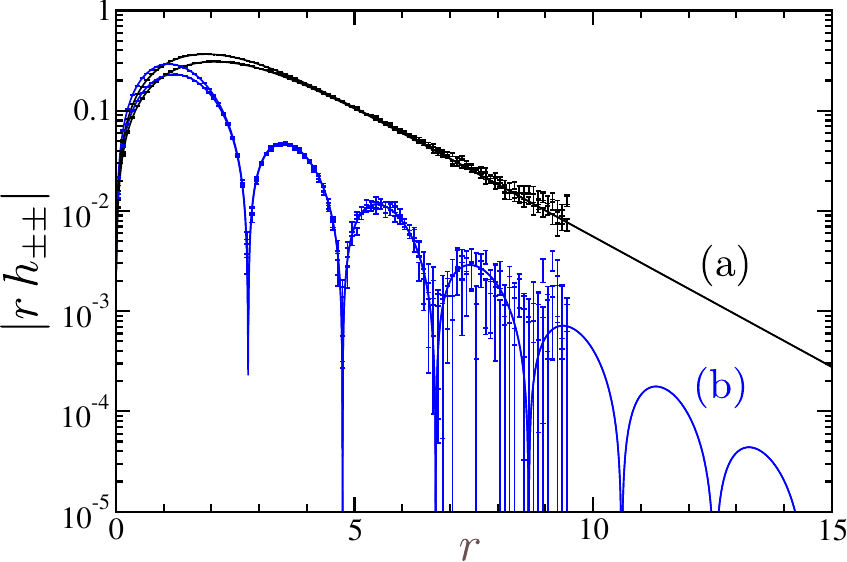}
\end{center}
\vskip -0.5cm
\caption{(color online) Pair distribution functions for the URPM at
  two state points on opposite sides of the Kirkwood line (see text
  for details), plotted as $|r\,h_{\pm\pm}|$ versus $r$ to illustrate
  the asymptotic behaviour.  Lines are HNC, data points with error
  bars are MC. Lengths are expressed in units of
  $\sigma$.\label{fig:decay}}
\end{figure}

\begin{figure}
\begin{center}
\includegraphics[clip=true,width=3.0in]{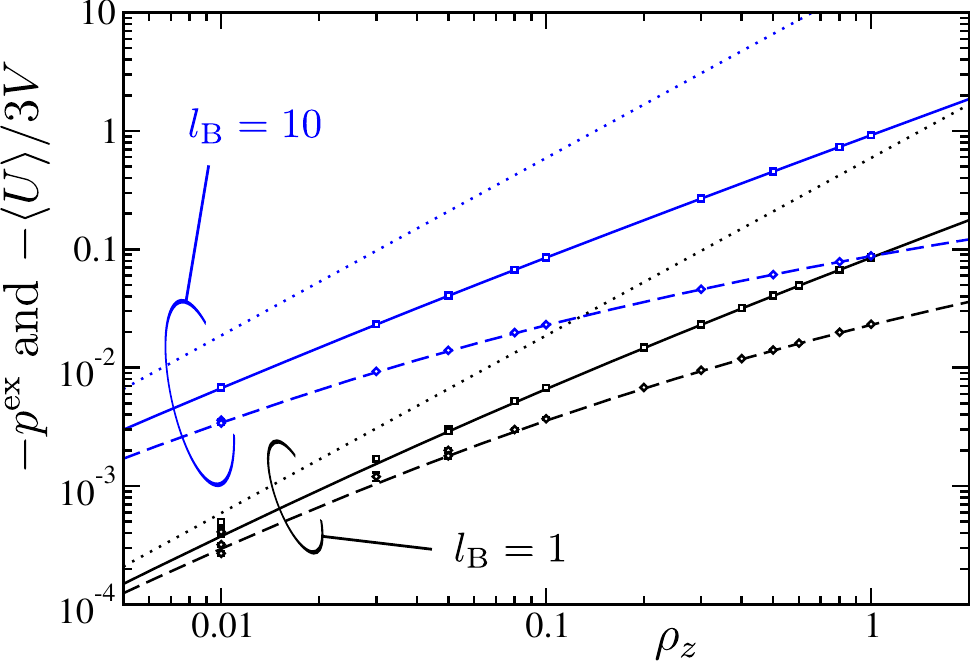}
\end{center}
\vskip -0.5cm
\caption{(color online) URPM thermodynamics along two isotherms
  showing $-\pex$ (solid HNC lines with square MC data points) and
  $-\myav{U}/3V$ (dashed HNC lines with diamond MC data points) as a
  function of density.  The Debye-H\"uckel limiting law is shown as a
  dotted line in the two cases.  Lengths and densities are expressed
  in units of $\sigma$, and thermodynamic quantities in units of
  $\kT/\sigma^3$.\label{fig:eos}}
\end{figure}

\begin{figure}
\begin{center}
\includegraphics[clip=true,width=3.0in]{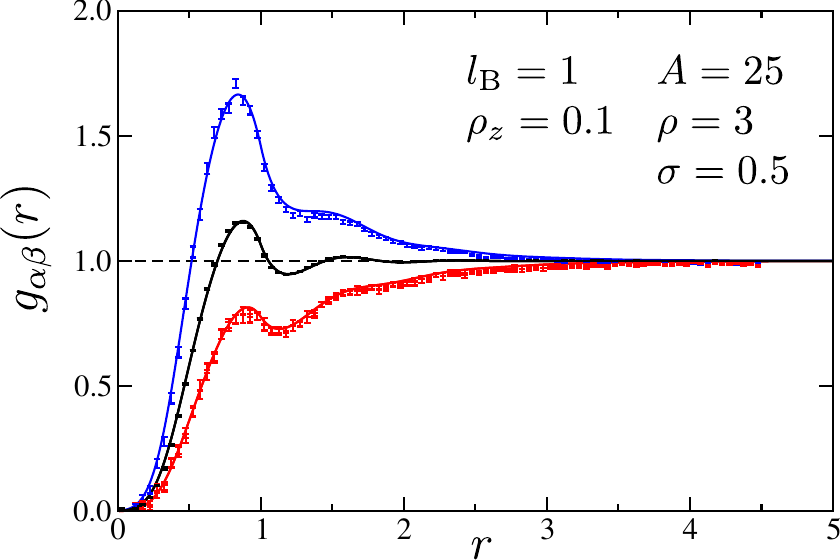}
\end{center}
\vskip -0.5cm
\caption{(color online) Pair distribution functions for a solvated
  model at the indicated state point.  Lines are HNC, data points with
  error bars are MC.  From top to bottom the curves are: $g_{+-}$;
  $g_{0+} = g_{0-}\approx g_{00}$; and $g_{++} = g_{--}$.  The
  difference between $g_{0\pm}$ and $g_{00}$ is tiny and not resolved
  in this plot. Lengths and densities are expressed in units of
  $r_c$.\label{fig:mcgr}}
\end{figure}

\section{Results}
\subsection{Comparison between HNC and MC}
We first consider the URPM as a baseline.  Fig.~\ref{fig:decay}
shows the pair distribution functions plotted as $|r\,h(r)|$ at two
state points on either side of the Kirkwood line: (a)
$\rhoz\sigma^3=0.02$ and $\lB/\sigma=1$ where the asymptotic decay is
purely exponential; and (b) $\rhoz\sigma^3=0.2$ and $\lB/\sigma=10$
where the decay is damped oscillatory.  The agreement between HNC and
MC is excellent.  These MC simulations are expensive to perform even
in the absence of a neutral solvent species.  For (a) and (b)
respectively, $10^6$ and $10^5$ MC configurations were required to
reduce the errors to an acceptable level at large $r$ (note that there
are ten times as many particles for the latter state point).  A box of
size $(20\sigma)^3$ is necessary to reach out to $r\approx9\sigma$.
Each state point required nearly 3000 hours of CPU time, in marked
contrast to the HNC solution which takes less than a second.  This is
the origin of the claim earlier that HNC can be up to ten million
times faster than MC.

MC simulation of the thermodynamics is much less demanding.
Fig.~\ref{fig:eos} shows the excess pressure ($\pex=p-\rho\kT$) and
internal energy for the URPM along isotherms at $\lB/\sigma=1$ and
10. There is excellent quantitative agreement between HNC and MC.  For
the most part these simulations were carried out in a box of size
$(10\sigma)^3$.  For $\lB/\sigma=1$ though we did check for finite
size effects at the lowest investigated density by increasing the box
size to $(15\sigma)^3$ and $(20\sigma)^3$.  The results are shown in
Fig.~\ref{fig:eos} as the multiple data points at
$\rhoz\sigma^3=0.01$, and indicate that finite size effects are small.

The excess pressure and the internal energy (divided by three) are
both expected to trend to the Debye-H\"uckel limiting law,
$\pex=\myav{U}/3V=-\kD^3/(24\pi)$, as the density decreases.  As can
be seen, the approach is rather slow.  Note that the relation
$\pex=\myav{U}/3V$ is a consequence of Clausius' virial theorem
applied to point particles interacting with the Coulomb
potential \cite{Cla70}.

In the presence of a neutral solvent the attainable MC accuracy is
much diminished, largely because of the need to equilibrate the
solvent particles. Fig.~\ref{fig:mcgr} shows an example of pair
distribution functions for a solvated model at a typical state point.
Again there is excellent agreement between HNC and MC (box size $(10 r_c)^3$).
In this plot note that symmetry enforces $g_{0+}=g_{0-}$ and
$g_{++}=g_{--}$.  The approximate symmetry $g_{0\pm}\approx g_{00}$ is
only very weakly broken in HNC (and not at all in the RPA) since it is
exactly true that $U_{0\pm}=U_{00}$.

To summarise the key result of this section: HNC accurately
reproduces MC for the parameter ranges of interest.  Thus we conclude
that we can place enough confidence in HNC to use it as a tool to
explore the properties of the model.

\begin{figure}
\begin{center}
\includegraphics[clip=true,width=3.0in]{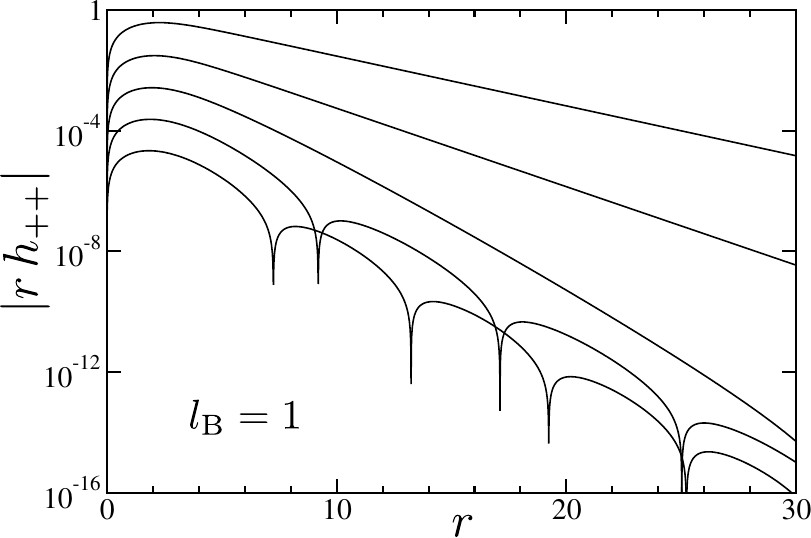}
\end{center}
\vskip -0.5cm
\caption{Total correlation function $h_{++}(r)$ from HNC for the URPM
  at $\lB=1$ and $\rhoz=0.01(1)5$ (top to bottom).  Curves have been
  displaced for clarity.  The Kirkwood transition from pure
  exponential decay ($\rhoz\alt0.03$) to damped oscillatory
  ($\rhoz\agt0.03$) is clearly seen. Lengths and densities are
  expressed in units of $\sigma$.\label{fig:scan}}
\end{figure}

\begin{figure}
\begin{center}
\includegraphics[clip=true,width=3.0in]{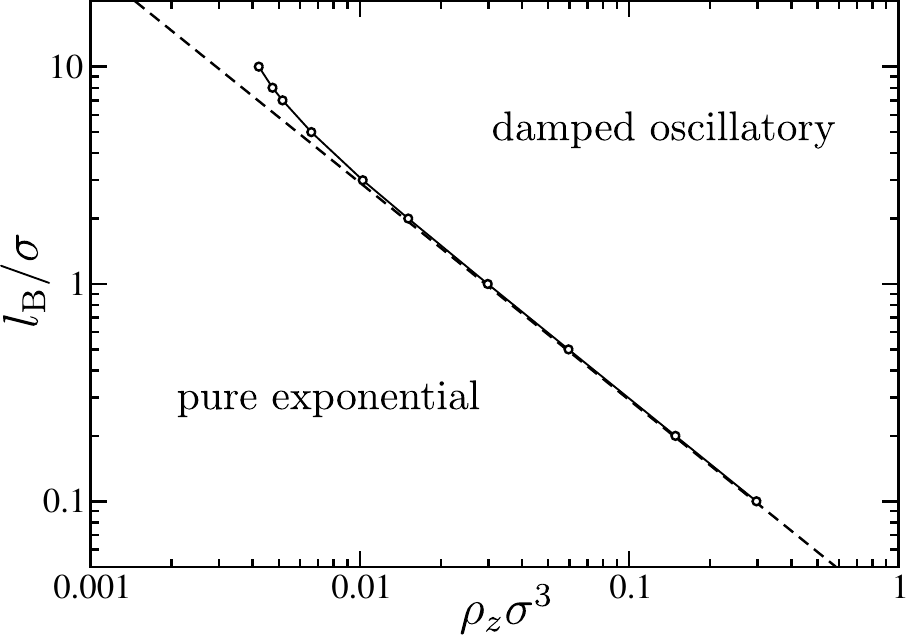}
\end{center}
\vskip -0.5cm
\caption{The Kirkwood line for the URPM.  The solid line with circles
  is from HNC.  The dashed line is the RPA, from Eq.~\eqref{eq:klrpa}.
  If solvent particles and short range repulsions are added, this map
  is practically unchanged.\label{fig:kirkwood}}
\end{figure}

\begin{figure}
\begin{center}
\includegraphics[clip=true,width=3.0in]{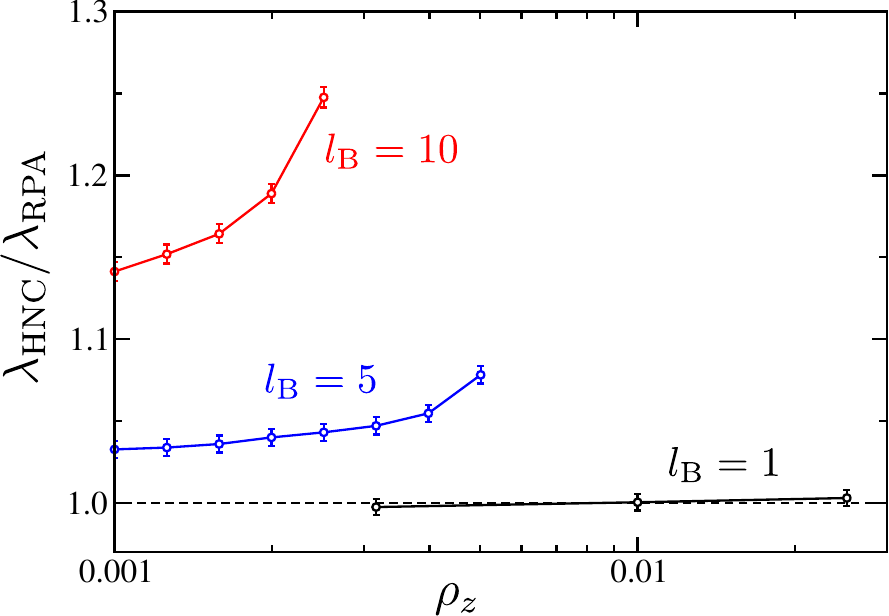}
\end{center}
\vskip -0.5cm
\caption{(color online) The screening length for the URPM, comparing
  the value extracted by fitting the asymptotic tails of
  $h_{\alpha\beta}$ in HNC, to the RPA value from
  Eq.~\eqref{eq:lrpa}. Lengths and densities are expressed in units of
  $\sigma$.\label{fig:lambda}}
\end{figure}

\begin{figure}
\begin{center}
\includegraphics[clip=true,width=3.0in]{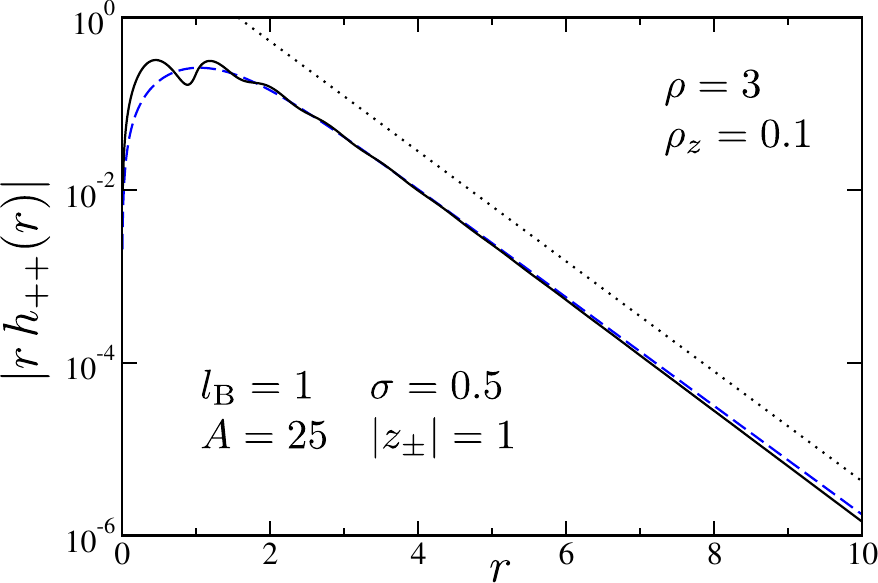}
\end{center}
\vskip -0.5cm
\caption{(color online) Comparison between a fully solvated model
  (solid line, black) and the URPM equivalent (dashed line, blue).
  Both are calculated using HNC.  The dotted line is the RPA
  prediction from Eq.~\eqref{eq:lrpa}.  Lengths and densities are
  expressed in units of $r_c$. \label{fig:solvent}}
\end{figure}

\subsection{Screening properties from HNC}
We now use the HNC to calculate the screening length from the
asymptotic behaviour of the computed total correlation functions,
focussing at first on the URPM.  Fig.~\ref{fig:scan} shows the typical
behaviour of $h_{++}(r)$ through the Kirkwood transition as $\rhoz$
varies at fixed $\lB$.  Note that the decay rate of the total
correlation function at first increases with increasing density, until
one crosses the Kirkwood line, after which the decay rate remains
roughly constant but the period of the oscillations decreases.  This
behaviour is similar to the RPA, and presumably reflects the pole
structure as discussed in section \ref{sec:rpasol}.

\begin{figure}
\begin{center}
\includegraphics[clip=true,width=3.0in]{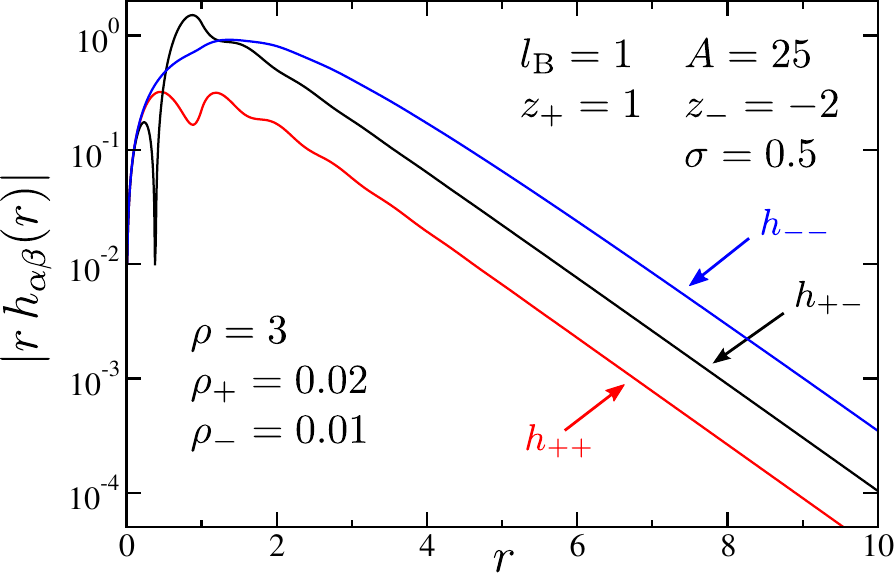}
\end{center}
\vskip -0.5cm
\caption{(color online) HNC results for a
  1:2 electrolyte.  Lengths and densities are expressed in units of
  $r_c$.\label{fig:asym}}
\end{figure}

One can `zero in' on the Kirkwood line transition by systematically
narrowing the range of densities which are plotted.  In this case one
finds the transition is located at
$\rhoz\sigma^3\approx(30.0\pm0.5)\times10^{-3}$.  By proceeding in
this way, the entire Kirkwood line can be mapped out in the
$(\rhoz\sigma^3,\lB/\sigma)$ plane.  This is shown in
Fig.~\ref{fig:kirkwood}, where HNC is compared to the RPA Kirkwood
line from Eq.~\eqref{eq:klrpa}.  We see that for $\lB/\sigma\alt5$ the
RPA is practically indistinguishable from HNC, and even at
$\lB/\sigma=10$ the difference is still less than 40\%.  Above this
value of $\lB/\sigma$ HNC ceases to converge to a solution.

On the low density side of the Kirkwood line the HNC screening length
can be found by fitting the tail of total correlation function to the
expected asymptotic behaviour in Eq.~\eqref{eq:ldef}.  Results along
isotherms at three values of $\lB/\sigma$ are shown in
Fig.~\ref{fig:lambda}.  Crucially, we see that for $\lB/\sigma\alt5$,
the RPA screening length from Eq.~\eqref{eq:lrpa} is in error by less
than 10\% compared to HNC.

All the results presented so far have been for the URPM in the absence
of a neutral solvent species.  Remarkably, we have found that very
little changes if short range repulsions are added ($A=25$) and a
solvent is included ($\rho r_c^3=3$).  For example the Kirkwood line
in Fig.~\ref{fig:kirkwood} is practically unchanged and we have found
the same to be true for the screening length itself.  We give a single
example here.  Fig.~\ref{fig:solvent} shows the asymptotic decay of
$h_{++}(r)$ for the indicated state point for a fully solvated model,
compared to the equivalent URPM at $\lB/\sigma=2$ and
$\rho_z\sigma^3=0.0125$.  A line indicating the RPA decay from
Eq.~\eqref{eq:lrpa} is also included.  We see that the presence of a
solvent and short range repulsions confers some liquid structure at
short distances but the asymptotic decay is practically unchanged, and
agrees well with the RPA.

Lastly we turn to the more complicated case of an asymmetric
electrolyte.  Fig.~\ref{fig:asym} shows the total correlation
functions for a 1:2 electrolyte calculated using HNC.  As can be seen
the asymmetry splits apart the three ionic correlation functions, but
nevertheless they share a common decay length, $\lHNC/r_c\approx
0.93$. This can be compared to $\lRPA/r_c\approx 1.02$, calculated
from Eq.~\eqref{eq:lrpa} using $\rho_zr_c^3=0.06$
(\ie\ $\rho_z=\rho_++4\rho_-$).  The difference between HNC and RPA is
less than 10\%, as might be expected from Fig.~\ref{fig:lambda} since
the 1:2 case is intermediate between the 1:1 case ($\lB/\sigma=2$) and
the 2:2 case ($\lB/\sigma=8$).

The main conclusions from this section are: first the solvent has
practically no effect on the screening properties so that
Fig.~\ref{fig:kirkwood} can be used as a quasi-universal quide, and
second for many applications, such as to 1:1 electrolytes, the RPA
suffices.

\begin{figure}
\begin{center}
\includegraphics[clip=true,width=3.0in]{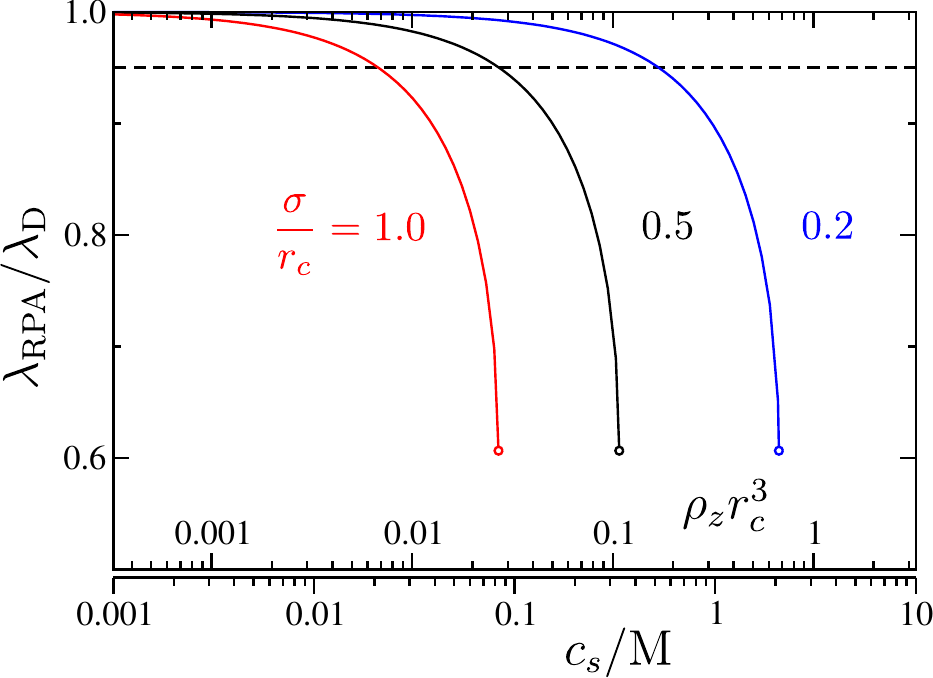}
\end{center}
\vskip -0.5cm
\caption{(color online) Ratio between RPA screening length and Debye
  length for a 1:1 electrolyte, as a function of concentration, for
  three choices of $\sigma$.  The lower (upper) horizontal axis shows
  the concentration in physical (simulation) units.  Each curve
  terminates when the model system crosses the Kirkwood line.  The
  dashed line is at $\lRPA/\lD=0.95$.\label{fig:csmap}}
\end{figure}

\begin{table}
\begin{ruledtabular}
\begin{tabular}{llccl}
%1\qquad & $c_s/\M$ & \multicolumn{2}{c}{\quad 0.1\phantom{00}} & \qquad \\
1 & $\lB/r_c$ & \multicolumn{2}{c}{1.09\phantom{0}}\\
2 & $\rho_zr_c^3$ & \multicolumn{2}{c}{0.032} \\
3 & $\lD/r_c$ & \multicolumn{2}{c}{1.50\phantom{0}}\\
\hline
4 & $\sigma/r_c$ & \phantom{$-$}0.5\phantom{00} & 
   \phantom{$-$}1.0\phantom{0} \\
5 & $\lB/\sigma$ & \phantom{$-$}2.17\phantom{0} & 
   \phantom{$-$}1.09 \\
6 & $4\pi e \lB\rho_z \sigma^2$ & \phantom{$-$}0.30\phantom{0} & 
   \phantom{$-$}1.20 \\
7 & $W_0(-4\pi \lB\rho_z \sigma^2)$ & $-0.125$ & $-0.88\pm0.60\,i$ \\
8 & $\lRPA/r_c$ &  \phantom{$-$}1.41\phantom{0} & \phantom{$-$}---\\
9 & $\lRPA/\lD$ & \phantom{$-$}0.94\phantom{0} & \phantom{$-$}---
\end{tabular}
\end{ruledtabular}
\caption[?]{Sample calculation for a $0.1\,\M$ 1:1 electrolyte.\label{tab:eg}}
\end{table}

\subsection{Worked example}
\label{sec:eg}
Let us work through the example given at the end of section
\ref{sec:ie}, for which the RPA solution is applicable.  The
calculations are shown in the numbered rows in Table \ref{tab:eg}.  We
start from the standard DPD mapping with $r_c=0.645\,\nm$ and
$\lB=0.7\,\nm$.  This gives $\lB/r_c=1.09$ (row 1).  If the molar
concentration of a 1:1 electrolyte is $c_s$, then $\rho_z=2\times
10^{3} c_s N_A$ (row 2; the factor two accounts for both species of
ion).  The Debye length (row 3) follows from Eq.~\eqref{eq:ld}, here
in the form $\lD=(4\pi\lB\rho_z)^{-1/2}$.  Alternatively one can use
the well known expression $\lD=0.31\,\nm/\sqrt{c_s}$ from the
colloidal literature \cite{VO48}. We choose a value for $\sigma$ (row
4) and calculate the left hand side of the inequality in
Eq.~\eqref{eq:klrpa} (row 6).  For the choice $\sigma=r_c/2$,
Eq.~\eqref{eq:klrpa} is satisfied and we are on the low density side
of the Kirkwood line.  We can then use the value of the Lambert
function (row 7) to calculate $\lRPA$ (row 8) from
Eq.~\eqref{eq:lrpa}.  Given that $\lB/\sigma\alt5$ (row 5), this
should be a good estimate of the true screening length. As claimed in
section \ref{sec:ie}, the answer deviates from the Debye length by
only 6\% (row 9).  For the choice $\sigma=r_c$ though,
Eq.~\eqref{eq:klrpa} is violated (row 6), and we are on the high
density side of the Kirkwood line.  This is also indicated by the fact
that the Lambert function (row 7) evaluates to a complex number.

\subsection{The choice of $\sigma$}
As we have seen, a mapping to a physical system fixes $\lB$ and $r_c$,
but the choice of $\sigma$ remains unresolved.  Our study so far
reveals this choice is a balance of conflicting requirements.  On the
one hand we would like to increase $\sigma$ as much as possible,
mainly because this reduces the cost of computing the electrostatic
interactions in a simulation.  On the other hand if $\sigma$ is too
large we run the risk of deviating strongly from the expected
screening properties of the physical system, and may ultimately cross
the Kirkwood line in Fig.~\ref{fig:kirkwood}.  Such behaviour is
almost certain to be artefactual since the chances of coinciding with
similar behaviour in the physical system seem remote.

For example for a 1:1 aqueous electrolyte we can plot the ratio
$\lRPA/\lD$ as a function of salt concentration $c_s$, using the
method just described in section \ref{sec:eg}.  Fig.~\ref{fig:csmap}
shows just such a plot, for three choices of $\sigma/r_c$.  Inspection
suggests as sensible compromise might be $\sigma/r_c=0.5$, since this
restricts significant deviations from the Debye length (\ie\ more than
10\%) to $c_s\agt 0.15\,\M$, where in any case the Debye length is
starting to become comparable to $r_c$.

\section{Discussion}
Let us close with some more remarks about implementation, and indicate
avenues for future work.  First, let us dispose of an elementary
point.  The simulations described here have been performed using MC,
rather than DPD.  The reason for this is that we are interested in
equilibrium properties, and MC is free from issues such as the choice
of integration algorithm and time step \cite{GW97}. Nevertheless the
Ewald method can easily be applied to a dynamical simulation, by
calculating the forces that arise from the potential energy in
Eq.~\eqref{eq:uewald}.

The usual Ewald implementation for point charges introduces a
`splitting parameter' so that part of the interaction is calculated in
real space and part in reciprocal space \cite{AT87}. Here we have
`physical-ised' the splitting parameter by linking it to the Gaussian
charge size $\sigma$, so that we can discard the real-space
interaction.  This may not always be the best choice, since one cannot
then optimise the splitting parameter to match the simulation box
size \cite{FS02}. However Coslovich \etal\ found that there is
practically no benefit in divorcing the splitting parameter from the
Gaussian charge size, at least for the URPM for their parameter
ranges.  Nevertheless it is worth bearing in mind this possibility,
particularly if $\sigma$ is much smaller than the simulation box size.

Aside from standard Ewald, any existing molecular dynamics (MD)
method could be used in principle to calculate electrostatic
interactions in DPD.  Most notable are the P3M
(particle-particle-particle-mesh) methods, such as that introduced by
Groot \cite{Gro03}, and hybrids such as smooth particle mesh
Ewald \cite{EPB+95}. Some of these methods are highly
parallelisable \cite{BLL98}, or highly efficient in other
ways \cite{EPB+95}. These MD methods are typically developed for point
charges, but the application to smeared charges should involve a
straightforward extension to the underlying algorithms.

As mentioned in the introduction, there is no consensus on the best
form of charge smearing (linear, exponential, Gaussian, \etc), nor,
perhaps, does there need to be.  All smearing methods generate pair
potentials which may differ in the short range part, but share a
common $\lB/r$ dependence for large $r$.  This raises the question of
whether the methods can be mapped on to one another, {\it vis \`a vis}
the screening properties.  Related to this is our observation that
short range repulsions have practically no effect on the screening
properties.  Whilst this is a great bonus for applications, it cannot
hold generally, for it would imply that there should be negligible
effect of the choice of smearing.  But we know this is manifestly
untrue: two Gaussian charge models with $\sigma'\ne\sigma$ do not have
the same screening properties.  More generally, in any smeared charge
model, another length scale must be present to non-dimensionalise
$\lB$.  The question of how to determine this length scale remains
unsolved.  Our present results suggest that a systematic use of HNC
could give an answer, thus providing a `Rosetta Stone' tying together
the existing treatments of DPD electrostatics.  This is the subject of
ongoing investigations.

Separate from this, a long term goal is to incorporate specific ion
effects into the model, such as the Hofmeister series \cite{CW85}.  As
mentioned in section \ref{sec:model}, we have here focussed on a
constant repulsion amplitude matrix $A_{ij}=A$.  Obviously there is
scope to go beyond this, using HNC to calculate both the structural
and thermodynamic consequences of unequal repulsion amplitudes.
The hope is that a suitable choice of $A_{ij}$ can be found, which
will systematically and transferrably capture specific ion effects.
It is encouraging to note in this respect that a similar programme has
been pursued with some success recently, for MD \cite{KD09, VLK+09}.

% \bibliography{scm,screening_v2}

\begin{thebibliography}{35}
\bibitem [{Wno()}]{Wnote}%
  \BibitemOpen
  \href@noop {} {}\bibinfo {note} {The Lambert function $W_0(x)$ can be found
  in commercial packages such as Mathematica and MATLAB, in open source
  packages such as Maxima and Octave, in the R statistical platform, and in the
  GNU Scientific Library. One-off evaluations can be performed on the internet
  using the Wolfram Alpha computational knowledge engine, \eg\ enter `{\tt
  LambertW[-0.11]}' for the $\sigma/r_c=0.5$ result in row 7 in Table
  \ref{tab:eg}.}\BibitemShut {Stop}%
\end{thebibliography}

%merlin.mbs apsrev4-1.bst 2010-07-25 4.21a (PWD, AO, DPC) hacked
%Control: key (0)
%Control: author (8) initials jnrlst
%Control: editor formatted (1) identically to author
%Control: production of article title (-1) disabled
%Control: page (0) single
%Control: year (1) truncated
%Control: production of eprint (0) enabled
%

\end{document}